\documentclass[twocolumn,showpacs,preprintnumbers,amsmath,amssymb,showkeys]{revtex4}

\draft 

\usepackage{graphicx}
\usepackage{dcolumn}
\usepackage{bm}

\begin{document}


\title{Fast detection of single-charge tunneling to a graphene quantum dot in a multi-level regime}

\author{T. M\"uller\footnote{Electronic address: thommuel@phys.ethz.ch}\footnote{Present address:
Department of Chemistry, University of Cambridge, Cambridge CB2 1EW,
UK}}
\author{J. G\"uttinger\footnote{Present address: CIN2 (ICN-CSIC), Catalan Institue of Nanotechnology, Campus de la UAB, 08193 Bellaterra (Barcelona), Spain}}
\author{D. Bischoff}
\author{S. Hellm\"uller}
\author{K. Ensslin}
\author{T. Ihn}
\affiliation{Solid State Physics Laboratory, ETH Z\"urich, 8093
Z\"urich, Switzerland}
\date{\today}

\begin{abstract}
\emph{In situ}-tunable radio-frequency charge detection is used for
the determination of the tunneling rates into and out of a graphene
single quantum dot connected to only one lead. An analytical model
for calculating these rates in the multi-level tunneling regime is
presented and found to correspond very well to our experimental
observations.
\end{abstract}

\maketitle

Time-resolved charge detection on quantum dots is a powerful
technique to straightforwardly extract single-particle transport
properties such as the occupation probability of quantum dots
determined by the Fermi distribution function in the leads and the
dot-lead tunneling rates~\cite{schleser:04,vandersypen:04}.
Recording the charge-detector signal through radio-frequency (rf)
reflection measurements~\cite{schoelkopf:98} can enhance the time
resolution drastically, enabling studies of systems with larger
tunneling rates.

To date, these types of experiments mainly study quantum dots in the
single-level regime, and investigations involving transport through
more than one level have focused on the phenomenon of
super-Poissonian noise~\cite{belzig:05,gustavsson:06b,fricke:07}.
Here, we present measurements and an analytical calculation of
multi-level tunneling rates using a graphene single quantum dot
connected to one lead.

\begin{figure}
\includegraphics{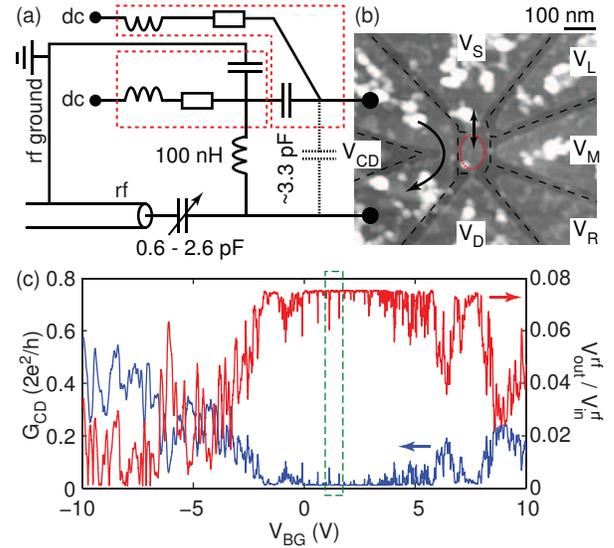}
\caption{\label{fig:setup}(Color online) (a) Schematic
representation of the circuitry used for rf and dc measurements. The
dotted black capacitor is the total stray capacitance, and the
dashed red boxes encase on-chip bias tees allowing for separation of
low- and high-frequency signals. (b) Atomic force microscopy image
of the graphene device studied in our experiments. Dashed black
lines mark regions of graphene etched away to form a
nanoconstriction at the location of the single-sided arrow and a
quantum dot encircled in dashed red. The nanoconstriction charge
detector is attached to the resonant circuit as a resistive element
$R$. (c) Conductance of the nanoconstriction measured via dc current
(blue trace, left axis) and rf reflectometry (red trace, right
axis).}
\end{figure}

Our charge-detection experiments were performed by incorporating a
graphene nanoconstriction charge detector, capacitively coupled to a
graphene single quantum dot, into an \emph{in situ}-tunable resonant
circuit~\cite{mueller:10}. A schematic view of the latter and an
atomic force microscopy image of the former are presented in
Figs.~\ref{fig:setup}(a) and (b), respectively. The structure was
fabricated through reactive ion etching on a mechanically exfoliated
graphene flake \cite{novoselov:04}. The parts of the graphene flake
etched away are emphasized by dashed black lines in
Fig.~\ref{fig:setup}(b) to enhance visibility. The roughly 100 nm
sized quantum dot (marked by the dashed red ellipse) and the
nanoconstriction (designated by the single-sided arrow) can be tuned
by in-plane gates and a global backgate. In spite of the symmetric
design of the barriers forming the quantum dot, only the upper lead
(source; represented by the double-sided arrow) was tunnel-coupled
to the dot. Therefore, we were not able to measure direct transport
through the graphene quantum dot. A more detailed description of the
fabrication procedure for this kind of devices can be found in
Ref.~\cite{guettinger:09}.

Due to the large stray capacitance formed between the metallic bond
pads and the highly doped silicon backgate, high-quality rf matching
of graphene samples using the standard circuit design forming a
resonance between a series inductor and the total stray capacitance
is difficult. One way to avoid this challenge is to remove the
backgate underneath the bond pads, which however complicates the
already involved fabrication of graphene nanostructures. We chose an
alternative circuit design consisting of a tunable capacitance in
series to a parallel $LCR$ circuit \cite{mueller:10}. Doing so and
harnessing the large change in conductance of the nanoconstriction
upon addition of a single charge to the quantum dot due to the
proximity of the structures ($\Delta G\approx0.08\times2e^2/h$ at
$G\sim0.1\times2e^2/h$) yields a very good charge sensitivity of
$\delta
q\approx3\times10^{-4}~e/\sqrt{\textrm{Hz}}$~\cite{vink:07,reilly:07,cassidy:07,mueller:10,barthel:10}.

All our measurements were performed in a variable-temperature insert
at a temperature of roughly 2~K.

The blue curve in Fig.~\ref{fig:setup}(c) shows the dc conductance
(left axis) through the nanoconstriction charge sensor as a function
of backgate voltage, while the red curve is the simultaneously
acquired rf reflection coefficient. Both measurements reveal a
transport gap in the backgate range between around $-1$ and 4~V.
Subsequent measurements are performed in the regime of the dashed
green box, in proximity of the charge-neutrality point of the
graphene charge sensor~\cite{Note2}.

\begin{figure}
\includegraphics{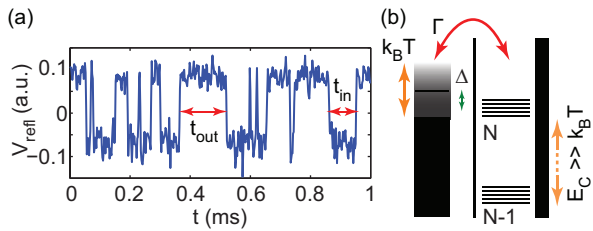}
\caption{\label{fig:Trace}(Color online) (a) Time-dependent rf
signal, 8-th order Bessel low-pass filtered at 200 kHz and sampled
at 500 kHz, revealing the times $t_\textrm{in}$ and $t_\textrm{out}$
the quantum dot is occupied or unoccupied by a single negative
excess charge. (b) Sketched energy diagram corresponding to the
multi-level regime studied here. A quantum dot with a closely-spaced
energy spectrum is connected to a lead with a tunnel rate $\Gamma$.
The Fermi function of the lead is softened by a temperature $T$
which is much larger than the level spacing $\Delta E$ but much
smaller than the charging energy $E_C$. The difference between the
Fermi level in the lead and the chemical potential of the quantum
dot is denoted by $\Delta$.}
\end{figure}

Tuning all gate voltages such that the quantum dot is at a
charge-degeneracy point between $N-1$ and $N$ negative charges on
the dot~\cite{Note3} and recording the charge-detector conductance
via rf reflectometry allows for studying tunneling of single charge
carriers in real-time, as shown in Fig.~\ref{fig:Trace}(a). This
trace has been 8-th order Bessel low-pass filtered at 200 kHz -
roughly two orders of magnitude faster than with previous
time-resolved experiments on graphene quantum
dots~\cite{guettinger:11} - and sampled at 500 kHz to ensure large
enough signal-to-noise ratio for reliable electron
counting~\cite{gustavsson:06}. From the average time a negative
charge spends inside (out of) the dot we can compute the dot
occupation probability and the rate for tunneling out of (into) the
quantum dot.

At our elevated temperatures we have good reason \cite{jacobsen:12}
to assume that we are in a multi-level tunneling regime - an
assumption that we will show to be correct in our discussion of the
experimental data. Thus we now calculate the tunneling rates of a
quantum dot connected to one lead in the multi-level regime
schematized in Fig.~\ref{fig:Trace}(b). Our analysis closely follows
the work of Beenakker~\cite{beenakker:91} in which all the
ingredients for the subsequent analysis are provided, and we also
adopt the notation thereof.

The multi-level regime of Coulomb blockade is characterized by a
temperature $k_\textrm{B}T$ much larger than the single-level
spacing $\Delta E$ but much smaller than the charging energy
$E_\textrm{C}$
\begin{equation}\label{Eq:MLTemp}
\Delta E\ll k_\textrm{B} T\ll E_\textrm{C}.
\end{equation}
The latter inequality is validated by the fact that we extracted an
approximate charging energy of $10-20~\textrm{meV}$ typical for
graphene quantum dots of this size (see for instance
Ref.~\cite{guettinger:11}), whereas we assume the former to be true
since we have not observed any sign of excited states in either the
similarly sized charge detector or the dot itself. Nevertheless, we
shall assert the inequality's validity later-on.

It has been shown in Ref.~\cite{beenakker:91} that the rate for
tunneling into a quantum dot occupied by $N-1$ charges is given by
the sum over all quantum dot states $p$ of their respective
couplings $\Gamma_p$ multiplied by the probability to have a state
at energy $E_p$ available in the lead (given by the Fermi
distribution function $f$) and the conditional probability
$1-F_\textrm{eq}(E_p|N-1)$ of having state $p$ empty in an
$N-1$-fold occupied dot in equilibrium
\begin{eqnarray}\label{Eq:GinSum}
   \nonumber \Gamma_\textrm{in}&=\sum_p & \Gamma_p\left[1-F_\textrm{eq}(E_p|N-1)\right]\\
   & &\times f(E_p+U_N-U_{N-1}-E_F),
\end{eqnarray}
where $U_N$ is the electrostatic energy of a dot containing $N$
negative charge carriers, and $E_F$ is the chemical potential in the
lead(s). Beenakker has also elucidated that $F_\textrm{eq}(E_p|N)$
can be expressed as a Fermi function $f(E_p-\mu_N)$ in the
high-temperature limit, where the chemical potential $\mu_N$ of the
dot is determined by
\begin{equation}\label{Eq:mu}
    \sum_{p=0}^\infty f(E_p-\mu_N)=N.
\end{equation}

By assuming the couplings $\Gamma_p$ to be equal and independent of
energy, we can replace the sum by an integral over energy times
density of states $\rho$ which we will also assume to be
energy-independent. Thus we can perform the integration which yields
\begin{eqnarray}\label{Eq:Gin}
    \nonumber\Gamma_\textrm{in} &=& \frac{\rho\Gamma\left[-U_N+U_{N-1}+E_F-\mu_{N-1}\right]}{1-\exp\left[(U_N-U_{N-1}-E_F+\mu_{N-1})/k_\textrm{B}
    T\right]} \\
    & \equiv & \rho\Gamma\frac{\Delta}{1-\exp(-\Delta/k_\textrm{B}
    T)},
\end{eqnarray}
where $\Delta$ is the energy detuning between the chemical
potentials in the lead and the dot (note that our definition of
$\Delta$ has an additional minus sign compared to the one of
Ref.~\cite{beenakker:91}). Analogously, we can compute
\begin{eqnarray}\label{Eq:Gout}
    \nonumber\Gamma_\textrm{out} & = \sum_p & \Gamma_p F_\textrm{eq}(E_p|N)\\
    \nonumber &  &\times\left[1-f(E_p+U_N-U_{N-1}-E_F)\right]\\
    & = \rho\Gamma &\frac{-\Delta}{1-\exp(\Delta/k_\textrm{B} T)}.
\end{eqnarray}
In this last equality we have made use of the fact that at our
relevant energies $\mu_{N-1}\approx\mu_N$.

Adding these two expressions leads to the inverse of the correlation
time of the random telegraph signal \cite{machlup:54} given by
\begin{eqnarray}\label{Eq:gamma}
    \gamma=\Gamma_\textrm{in}+\Gamma_\textrm{out} = \Gamma\rho\Delta\frac{\sinh(\Delta/k_\textrm{B} T)}{\cosh(\Delta/k_\textrm{B}
    T)-1}.
\end{eqnarray}
We want to emphasize that in contrast to the above expression, the
sum of tunnel rates is constant with respect to detuning in the
single-level regime if the couplings for tunneling-in and -out are
equal (and are only varying over a range of the order of
$k_\textrm{B}T$ otherwise). Therefore, a strongly varying sum of
tunneling rates excludes a single-level regime.

The expected rate of events $r_E$ (the number of times per second an
electron tunnels in and out of the quantum dot) can now be
calculated to be
\begin{eqnarray}\label{Eq:reML}
    \nonumber r_E=\frac{\Gamma_\textrm{in}\Gamma_\textrm{out}}{\Gamma_\textrm{in}+\Gamma_\textrm{out}} & = & \frac{\Gamma\rho\Delta}{2\sinh(\Delta/k_\textrm{B}
    T)}\\
    & \approx & \frac{\Gamma\rho}{2} k_\textrm{B} T\cosh^{-2}\left(\frac{\Delta}{2.5 k_\textrm{B}
    T}\right),
\end{eqnarray}
equivalent to what Beenakker found for the conductance through a dot
in the multi-level regime (and what Kulik and Shekhter found for
transport through granular metals~\cite{kulik:75}) since in all
instances, tunneling-in and -out are treated on an equal footing.

\begin{figure}
\includegraphics{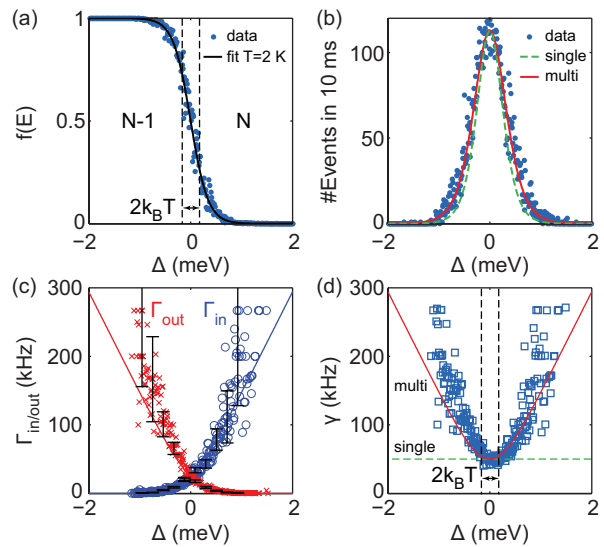}
\caption{\label{fig:ML} (Color online) (a) Electronic occupancy of
the lead as a function of dot-lead energy detuning $\Delta$. Every
blue data point is the fraction of time the dot is occupied by $N-1$
(instead of $N$) negative charges as extracted from traces as shown
in Fig.~\ref{fig:Trace}(a). The solid black line is a fit to a Fermi
distribution function in the lead, yielding a temperature of
$T=1.97\pm0.03~\textrm{K}$. The dashed black lines indicate the
magnitude of temperature. (b) Number of tunneling events in a 10 ms
long trace. The dashed green and solid red lines are calculated
rates for a single and multi-level regime, respectively, using the
temperature extracted in (a) and the height of the data points. (c)
Tunneling rates determined through the average dwell times in (red
crosses) and out (blue circles) of the quantum dot. The solid lines
are calculations using Eqs.~\ref{Eq:Gin} and \ref{Eq:Gout} in the
main text. For sake of clarity, the statistical error bars are only
shown for one in 20 traces. (d) Measured (blue squares) and
analytically determined (Eq.~\ref{Eq:gamma}; solid red line) sum of
the multi-level tunnel rates shown in (c). The dotted green line
indicates the behavior in the single-level regime.}
\end{figure}

Equipped with these analytical expressions for the tunneling rates,
we turn to our experimental results. Figure \ref{fig:ML}(a) shows
the fraction of time the quantum dot is devoid of the $N$-th excess
charge when sweeping over the $\{N-1,N\}$ charge-degeneracy point,
reflecting the Fermi distribution function of the lead. Fitting the
measured data (blue dots) with a Fermi function (solid black curve)
results in a temperature of $T=1.97\pm0.03~\textrm{K}$. Using this
temperature, we can compare the number of tunneling events extracted
from the same time traces with the expected number in a single-
(dashed green line; see Ref.~\cite{beenakker:91}) and multi-level
(solid red line; Eq.~\ref{Eq:reML}) regime in Fig.~\ref{fig:ML}(b).
We find that the solid red line corresponding to multi-level
tunneling matches the measurements better~\cite{Note4}.

In Fig.~\ref{fig:ML}(c) we show the experimentally extracted rates
for tunneling-in (blue circles) and tunneling-out (red crosses).
Additionally, we plot the analytical functions for the rates given
by Eqs.~\ref{Eq:Gin} and \ref{Eq:Gout} as solid lines. The only fit
parameter for these curves is the product $\Gamma\rho$ of tunnel
coupling and density of states which was chosen such that the
experimental results coincide with the calculations at zero
detuning. The error bars indicating the uncertainty of the
determination of the tunneling rates from the dwell
times~\cite{gustavsson:06b} are only shown for one in 20 data points
to enhance visibility. Finally, the sum of these tunneling rates is
presented in Fig.~\ref{fig:ML}(d) for both, experiment (blue
squares) and single- (dotted green line) as well as multi-level
(solid red line; Eq.~\ref{Eq:gamma}) models. As we observe a linear
increase in the sum of the rates for detunings above
$k_\textrm{B}T$, we can be certain to be in a multi-level regime.

In all plots of Fig.~\ref{fig:ML}, the horizontal axis has been
converted to energy using a lever arm set via the relative influence
of the swept backgate and the dot source lead~\cite{Note1}.

The qualitative and quantitative agreement between our experiments
and theoretical calculations is striking, especially considering
that only one free parameter $\Gamma\rho$ is available, since
temperature has been fixed by the Fermi fit in (a). Nevertheless, we
observe a deviation from the model for large detuning, where the
experimentally obtained values lie distinctly above our
expectations. This may originate from the fact that, in contrast to
our assumptions, either the density of states $\rho$ or the tunnel
couplings $\Gamma_p$ in the dot are not entirely independent of
energy. Furthermore, the couplings may be different for each state
$p$. Also, it is conceivable that only a finite number of states
contribute to tunneling. From the energy-dependence of the sum of
tunneling rates in Fig.~\ref{fig:ML}(d), though, we can definitely
exclude single-level tunneling. Missed events due to our large yet
finite bandwidth lead to a further enhancement of the numeric value
of the measured tunnel rates for large $\Delta$ and can therefore
not explain the deviations.

In conclusion we have performed time-resolved charge detection
measurements on a graphene quantum dot using rf reflectometry with
large bandwidth. From these experiments we have extracted the rates
for tunneling-in and tunneling-out of the dot, which are in good
agreement with our analytical model for multi-level tunneling to a
single lead. This type of fast charge detection on graphene quantum
dots opens possibilities for studying interactions of Dirac Fermions
in the framework of full counting statistics and is a great
instrument for determining spin relaxation times in graphene quantum
dot circuits.

We thank C. Barengo for invaluable technical assistance and B.
K\"ung and Y. Komijani for fruitful discussions. Funding by the
Swiss National Science Foundation (SNF) via National Center of
Competence in Research (NCCR) Nanoscale Science is gratefully
acknowledged.


\end{document}